\documentclass[aps,prl,twocolumn,showpacs]{revtex4-1}
\usepackage{graphicx} 
\usepackage{accents}
\usepackage{amssymb}
\usepackage{float}
\usepackage{gensymb}
\usepackage[title]{appendix}
\begin{document}
\title
{\bf Interaction Approach to Anomalous Spin Texture in Warped Topological Insulators}
\author{T. Hakio\u{g}lu}
\affiliation{
{Energy Institute and the Department of Physics, \.{I}stanbul Technical University, Maslak 34469, \.{I}stanbul, Turkey}
}
\begin{abstract}
Interactions between the surface and the bulk in a strong topological insulator (TI) cause a finite lifetime of the topological surface states (TSS) as shown by the recent experiments. On the other hand, interactions also induce unitary processes, which, in the presence of anisotropy and the spin-orbit coupling (SOC), can induce non-trivial effects on the spin texture. Recently, such effects were observed experimentally in the $Bi_2X_3$ family, raising the question that the hexagonal warping (HW) may be linked with new spin-related anomalies. The most remarkable among which is the 6-fold periodic canting of the in-plane spin vector. Here, we show that, this anomaly is the result  of a {\it triple cooperation between the interactions, the SOC and the HW}. To demonstrate it, we formulate the spin-off-diagonal self energy. A unitary phase with an even symmetry develops in the latter and modulates the spin-1/2 vortex when the Fermi surface is anisotropic. When the anisotropy is provided by the HW, a 6-fold in-plane spin-canting is observed. Our theory suggests that the spin-canting anomaly in $Bi_2X_3$ is a strong evidence of the interactions. High precision analysis of the spin-texture is a promising support for further understanding of the interactions in TIs.    
\end{abstract}
\pacs{71.70.Ej,/2.25-b,75.70.Tj}
\maketitle

The TIs have metallic, Dirac-like surface states and spin-momentum locking while the bulk remains as insulator with an inverted gap\cite{TI_general_intro}. Due to the time reversal symmetry (TRS) and the strong SOC, the intraband carrier backscattering is strongly suppressed. As a result, the single particle picture (SPP) is highly accurate in representing the electronic properties of the TSS\cite{topo_protect}. This makes the TIs a great promise for new generation dissipationless devices. However, recent experiments reveal the existence of scattering mechanisms between the TSSs and the bulk which clearly point out to deviations from the ideal SPP\cite{Park,Zhu,Pan,Valla,Barriga,Chen}. 
 
The interactions between the TSS and the bulk introduce finite quasiparticle lifetime on the surface. The optical phonon coupling is the main contribution as shown experimentally\cite{Zhu,Pan} and by ab-initio calculations\cite{Heid}. Spin-dependent interactions can also impair the spin-momentum locking and degrade the quality of the surface metallic state. Although the details of the interactions are yet to be studied microscopically, the experiments indicate a weak broadening of the surface density-of-states (DOS) pointing out that the TSS are long-lived.  
 
The 3D TIs of which the $Bi_2X_3, X=Se,Te$ family is of primary importance\cite{BiSe_exp,BiTe}, belong to the $D_{3d}$ point group of the rhombohedral family. The unit cell has the quintuple layered structure as $X-Bi-X-Bi-X$ coordinated hexagonally within each layer. The 2D nature of the electronic bands has been shown in the ARPES measurements\cite{BiSe_ARPES}. The spin-orbit coupling (SOC) ${\vec g}_{\vec k}$, has the in-plane component ${\vec g}_{xy\vec k}=(g_x,g_y,0)=\alpha (-k_y,k_x,0)$ where $\alpha$ is the SOC constant and ${\vec k}=(k_x,k_y)=k(\cos\phi,\sin\phi)$ is the 2D momentum of the TSSs. The out-of-plane component ${\vec g}_{z\vec k}=(0,0,g_z)$ is the cubic-Dresselhaus SOC\cite{Fu_1} $g_z=\lambda (k_+^3+k_-^3)/2$ where $k_\pm=k_x\pm i k_y$ and $\lambda$ is the coupling constant. The $g_{z\vec k}$ lowers the continuous rotational symmetry of the Fermi surface by a hexagonal anisotropic warping\cite{BiSe_exp,BiTe}.

In Ref. \cite{gedik}, circular dichroism ARPES revealed several key features of the spin in the warped Dirac cone in  $Bi_2Se_3$\cite{criticism} namely: a $2\pi/3$-periodic out of plane spin and a singly periodic in-plane spin anomalously canted with a $2\pi/6$-periodic modulation. This in-plane spin canting anomaly (SCA) is defined as an anomalous non-orthogonality of the in-plane spin to the momentum, i.e. ${\vec k}$, i.e. ${\vec S}_{xy \vec k}=S_{\perp \vec k}\,\hat{g}_{\vec k}+S_{\parallel \vec k}\,\hat{k}$ with $S_{\parallel \vec k} \ne 0$. It disappears along the high symmetry directions $\Gamma M$ and $\Gamma K$. Interestingly, similar effects were observed also in the Rashba-split metals (eg. Bi/Ag (111)) which show similar HW rather than circular shape\cite{DOI}.    

The authors of Ref.\cite{Basak} proposed a model for the SCA by adding to the warped Fermi surface a new fifth-order term in momentum. Their motivation is to reproduce the key features of the non-orthogonal state and its 6-fold symmetry. The proposed quintic term, which has a strength independent from the better-understood HW, has not been examined before theoretically or observed experimentally, and its contribution survives in the absence of HW. We take this as our motivation here and, using an approach beyond the SPP, devise a mechanism for the SCA based on the interactions and the HW. 
 
On the first hand, the backscattering within the topological surface bands is prohibited due to the time reversal symmetry and the spin-momentum locking. On the other hand, it is known experimentally that the surface carriers are exposed to many-body interactions. Primary ones are the electron-hole excitations, incoherent phonon emission/absorption and the scattering by non-magnetic impurities\cite{Park}. The ab-initio calculations reveal further evidence for the existence of weak electron-optical phonon interaction\cite{Heid}. Spin-dependent interactions were also proposed\cite{Barriga}. At low energies the dominant mechanisms are the phonon and impurity scattering and they manifest Lorentzian line-shapes in the imaginary part of the spin-diagonal self-energy (SE) $\Sigma_0({\vec k},\omega)$.   
  
Secondly, there is experimental evidence that the interactions are affected by the HW as shown by the anisotropic scattering rates\cite{Barriga} and the quasi-particle interference\cite{qp_interference}. It is therefore quite natural to expect that, the spin-off-diagonal component of the SE, which is directly related to the renormalized spin-texture, is also affected by the HW. It will be shown below that, this component produces the spin-texture anomalies. 
 
\begin{figure}[h]
~~~~~~~~~~~\includegraphics [height=6.9cm,width=9cm]{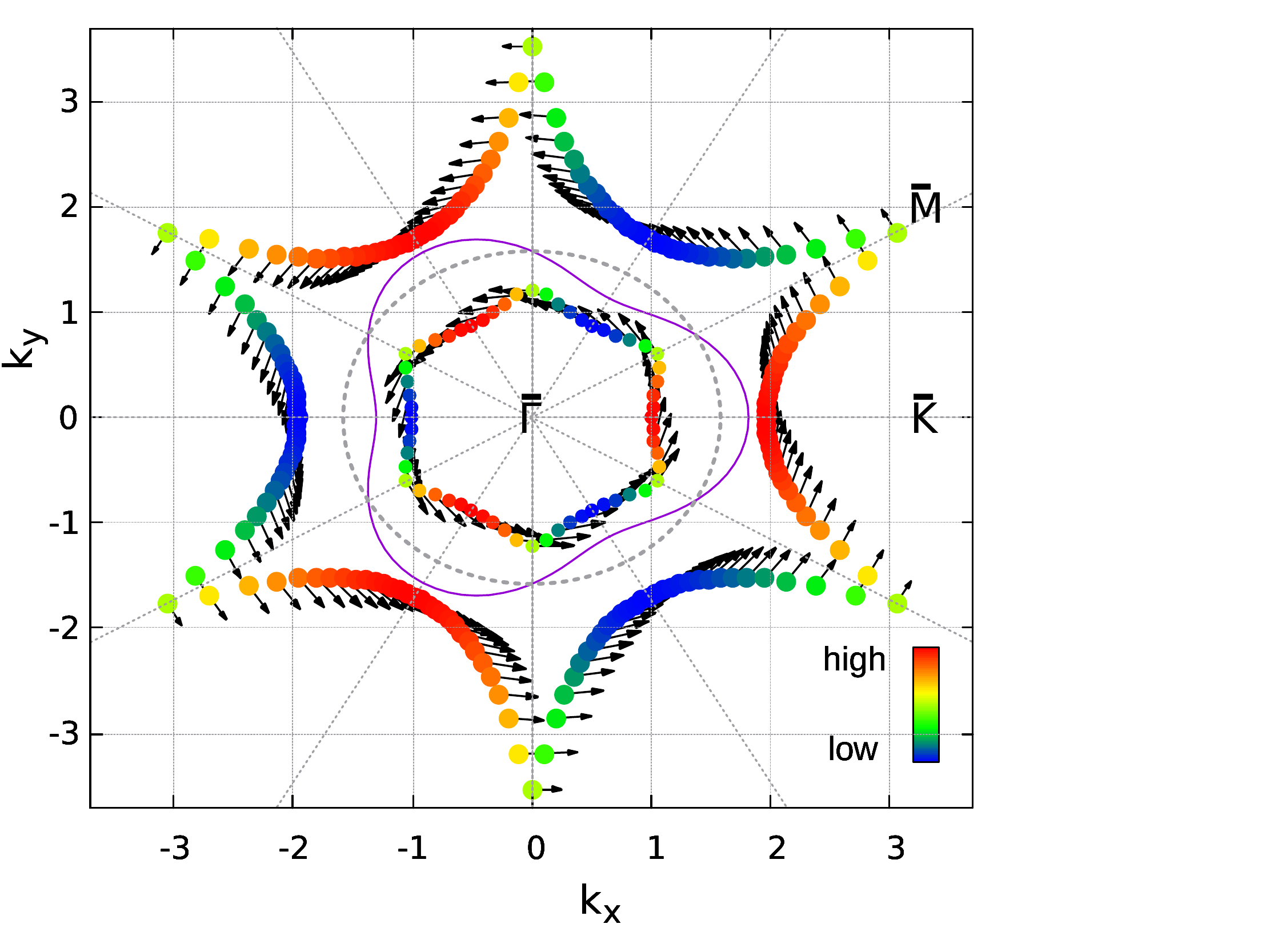} \\   
\vskip-0.3truecm
\caption{(Color online) The ${\vec S}_{\vec k}^{(+)}$ in Eq.(\ref{3}) is shown for two different energy levels. The in-plane component (depicted by the black vectors and normalized by the in-plane magnitude) is perpendicular to $\hat{k}$ independently from HW. The out-of-plane component is encoded in color scale. The $k_x,k_y$ are normalized by $d=\beta^{-1/4}\simeq 16 \AA$. Here "high/low" implies that the spin is fully out-of-plane "up/down". The 3-fold symmetric closed-line and the dotted circle are aid to the eye in the printed version.}   
\vskip-0.3truecm
\label{no-canting}
\end{figure}

Our aim here is to device a weakly interacting theory of the SCA without proposing a specific interaction mechanism. The reader will be motivated that, the present theory, in the light of new experiments, can reveal further information about the nature of the interactions. The bare Hamiltonian is 
\begin{eqnarray}
{\cal H}_0=E_{0{\vec k}}+{\vec g}_{\vec k}.{\vec \sigma}
\label{1}
\end{eqnarray}
where $E_{0{\vec k}}=\hbar^2 k^2/(2m)-\mu$ with $m$ as the electron band mass in the parabolic sector, $\mu$ is the chemical potential and ${\vec g}_{\vec k}={\vec g}_{xy\vec k}+{\vec g}_{z\vec k}$ is the total SOC with ${\vec g}_{xy\vec k}=\alpha k \hat{e}_\phi$ and ${\vec g}_{z\vec k}=\lambda k^3 \cos(3\phi) \hat{e}_z$. Here $\hat{e}_\phi$ and $\hat{e}_z$ are the unit vectors along the azymuthal and $z$ directions. The Eq.(\ref{1}) is diagonalized by the chiral basis  
$\vert \nu {\vec k}\rangle=\cos\frac{\theta_{\vec k}}{2} \vert \uparrow {\vec k}\rangle+\nu \frac{g_{+\vec k}}{g_{\vec k}} \,\sin\frac{\theta_{\vec k}}{2} \vert \downarrow {\vec k}\rangle$. Here $\nu=\pm$ is the branch chirality index, $g_{xy\vec k}=g_{x\vec k}+ig_{y\vec k}=\alpha k e^{i\phi_{S}}$, $g_{\vec k}=\vert {\vec g}_{\vec k}\vert=\sqrt{g_{x\vec k}^2+g_{y\vec k}^2+g_{z\vec k}^2}$, $\phi_{S}=-(\phi+\pi/2)$, $\cos\theta_{\vec k}=\lambda k^3 \cos(3\phi)/g_{\vec k}$. The $\vert \uparrow {\vec k}\rangle$ and $\vert \uparrow {\vec k}\rangle$ are the basis vectors of Eq.(\ref{1}) with spin quantization along the $z$-direction. The spin is given along the unit SOC vector $\hat{g}_{\vec k}={\vec g}_{\vec k}/g_{\vec k}$ as 
\begin{eqnarray}
{\vec S}_{\vec k}^{(+)}=\frac{\hbar}{2}{\hat g}_{\vec k} ~, \qquad {\vec S}_{\vec k}^{(-)}=-{\vec S}_{\vec k}^{(+)}~.
\label{3}
\end{eqnarray}
The spin ${\vec S}_{\vec k}^{(+)}$ in the $+$ band is depicted in Fig.\ref{no-canting} along the constant energy contours. 
In the well-studied SPP in Eq.'s(\ref{1}) and (\ref{3}), the in-plane spin direction, as shown by the arrows, is orthogonal to ${\vec k}$ independently from HW. The out-of-plane component has the three-fold symmetry. The predictions of the hexagonally warped SPP in Eq.(\ref{1}) for the out-of-plane spin\cite{Fu_1} agrees along the $\Gamma M$ direction, with the experimental results of Ref.\cite{gedik}.          

\begin{figure}[h]
\vskip-0.3truecm
\includegraphics[height=6.12cm,width=8cm]{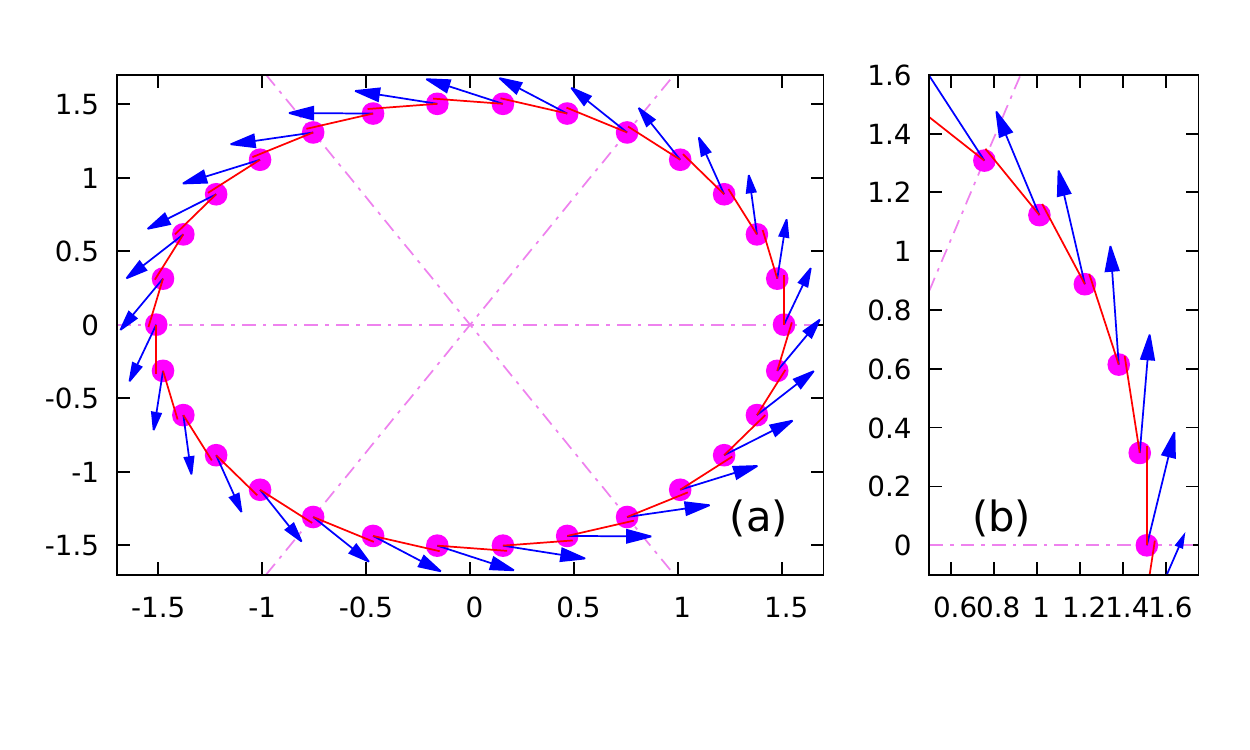}   
\vskip-1truecm
\caption{(Color online) a) The in-plane spin in Eq.(\ref{6b}) with $\tau_{mk}=0, m\neq 0$ (red lines are tangent to the circle at the dot positions), b) the spin is magnified between the equivalent $\Gamma K$ directions in (a). The axes are not shown for a larger view. The horizontal (vertical) axis is $k_x (k_y)$ in both figures.}  
\label{spin-canting_a}
\vskip-0.1truecm
\end{figure}

\begin{figure}[h]
\vskip-0.3truecm
\includegraphics [height=6.12cm,width=8cm]{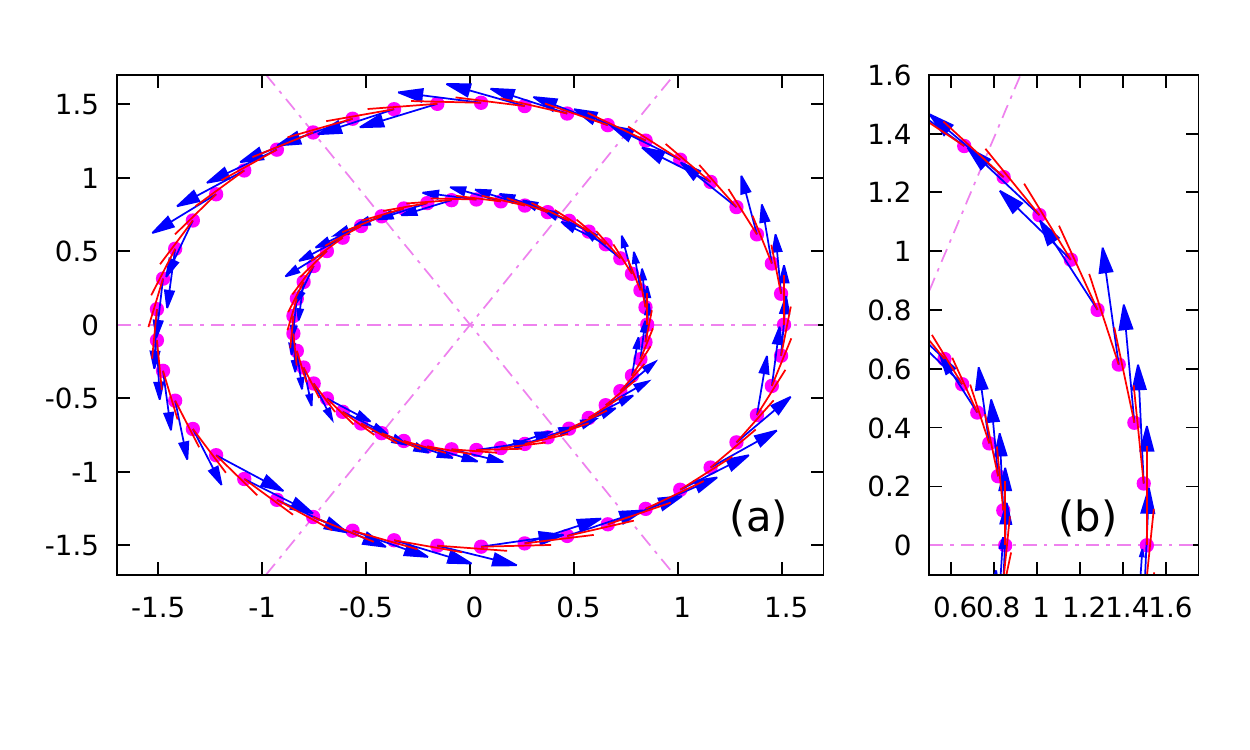}    
\vskip-1truecm
\caption{(Color online) a) The in-plane spin in Eq.(\ref{6c}) with $\tau_{mk}=0, m\neq 1$ with the six-fold symmetry (red lines are tangent to the circle at the dot positions), 
b) a single period of the spin vector oscillations magnified between the  $\Gamma K$ directions in (a). The axis labels are the same as in Fig.(\ref{spin-canting_a}).}  
\label{spin-canting_b}
\vskip-0.3truecm
\end{figure}

We turn our attention to the in-plane spin. In order to study the SCA we consider the effective isotropic and spin-independent interaction\cite{Park,Zhu,Pan,Valla,Barriga,Chen} $V_{\vert {\vec k}-{\vec k}^\prime\vert}$ for the electrons in the topological band. The SE in now described by $\Sigma({\vec k},\omega)={\Sigma}_0({\vec k},\omega)+{\vec \Sigma}({\vec k},\omega).{\vec \sigma}$ where the ${\vec \Sigma}({\vec k},\omega)$ describes the spin-dependent part. In the calculation of the SE, we will ignore the frequency dependent retardations and study the limit $\omega \to 0$ in $\Sigma({\vec k},\omega)$, i.e. $\Sigma_{\vec k}$. The renormalized Hamiltonian for the TSS then becomes
\begin{eqnarray}
{\cal H}^\prime&=&{\cal H}_0+\Sigma_{\vec k}\,, \qquad {\rm where} \quad \Sigma_{\vec k}=\Sigma_{0\vec k}+{\vec \Sigma}_{\vec k}.{\sigma}
\label{4} \\
{\vec \Sigma}_{\vec k}.{\sigma}&=&\pmatrix{\Sigma_{z\vec k} &  \Sigma_{xy\vec k} \cr 
{\Sigma^*_{xy\vec k}} & -\Sigma_{z\vec k}}\,,\quad \Sigma_{xy\vec k}=\Sigma_{x\vec k}-i\Sigma_{y\vec k}
\label{5}
\end{eqnarray} 
with $\Sigma_{0\vec k}$ as the spin independent part. Since the $\Sigma_{\vec k}$ is invariant under TRS, the $\Sigma_{0\vec k}$ is even and ${\vec \Sigma}_{\vec k}$ is odd. We ignore the $\Sigma_{z\vec k}$ here since it has no major role in the in-plane SCA. Also note that, an effective SOC can be defined as ${\vec G}_{\vec k}={\vec g}_{\vec k}+{\vec \Sigma}_{\vec k}$ in ${\cal H}^\prime$. We now write   
\begin{eqnarray}
\Sigma_{xy\vec k}=T_{\vec k}\,e^{i\phi_{S}} ~.  
\label{6}
\end{eqnarray} 
where, by the TRS, $T_{\vec k}$ is most generally a complex even function. Before tackling thes problem, it is instructive to recall the isotropic limit in the Eq.(\ref{6}). This is realized by the absence of HW where the $T_{\vec k}$ is isotropic and real. Recalling that, the Eq.(\ref{3}) is now replaced by 
\begin{eqnarray}
{\vec S}_{\vec k}^{(+)}=\frac{\hbar}{2}{\hat G}_{\vec k} ~,\quad {\hat G}_{\vec k}=\frac{{\vec G}_{\vec k}}{\vert {\vec G}_{\vec k} \vert}
\label{3prime}
\end{eqnarray}
the spin is then parallel to the initial ${\hat g}_{\vec k}$ and no SCA is observed. From here on, we confine to the upper energy band and drop the band index. It is necessary to observe that, away from the isotropic limit, a complex $T_{\vec k}$ is allowed in Eq.(\ref{6}) by unitarity, i.e. $T_{\vec k}=t_{\vec k} e^{i\Lambda_{\vec k}}$ with $t_{\vec k}$ and $\Lambda_{\vec k}$ real an even. Using Eq.(\ref{6}), we then have ${\vec \Sigma}_{\vec k}=Re\{T_{\vec k}\} \hat{g}_{xy\vec k}+ Im\{T_{\vec k}\} \hat{k}$. Here $\hat{g}_{xy\vec k}$ is the unit vector along the in-plane SOC and $\hat{k}={\vec k}/k$. This demonstrates that $T_{\vec k}$ must acquire an imaginary part to violate the spin-momentum orthogonality. 
The experimental SCA angle $\delta_{\vec k}$ in \cite{gedik} is then related to the $Im\{T_{\vec k}\}$ by $\sin \delta_{\vec k}=\hat{G}_{\vec k}.{\hat k}=Im\{T_{\vec k}\}/\vert {\vec G}_{\vec k}\vert$. For small angles, 
\begin{eqnarray}
\delta_{\vec k}\simeq \hat{G}_{\vec k}.\hat{k} \simeq \frac{t_{\vec k}}{\vert {\vec G}_{\vec k}\vert}\,\sin\Lambda_{\vec k}
\label{7.a} 
\end{eqnarray} 
We now examine the $T_{\vec k}$. Since the full SE matrix $\Sigma_{\vec k}$ in Eq.(\ref{4}) is restricted by the $C_{3v}$ symmetry, the $T_{\vec k}$ must have at least 6-fold symmetry expressed by 
\begin{eqnarray}
T_{\vec k}=\sum_{m=-\infty}^\infty\, T_{mk} e^{i6m\phi} 
\label{6a}
\end{eqnarray}
The $T_{0k}$ in Eq.(\ref{6a}) is the leading term when HW is weak. The other coefficients $T_{mk}$ generate an undulating pattern for the spin with $6m$-folded symmetry. On one hand, the nature of the interaction is imprinted in the full set of $T_{mk}$'s. On the other hand, they can be used to connect to the experiment. To show that, we define the $m$'th symmetry moment as $\tau_{mk}=(T_{mk}-T^*_{-mk})/2i$. From Eq.'s(\ref{3prime})-(\ref{6a}), $\tau_{mk}=\langle e^{-i6m\phi} {\vec S}_{\vec k}.\hat{k}\rangle_\phi$ where $\langle \dots \rangle_\phi$ denotes the angular average over $\phi$. For large $m$, the $\tau_{mk}$ is expected to get weaker. Here we will only examine the $m=0, 1$. For $m=0$, if 
\begin{eqnarray}
\tau_{0k}=\frac{2}{\hbar}
\langle {\vec S}_{\vec k}.\hat{k}\rangle_\phi
\label{warping_vs_canting2}
\end{eqnarray}
is nonzero, the spin rotates around the $2\pi$ range asymmetrically (with an outward or an inward bias). In the extreme case $\tau_{0k}\ne 0 ; \tau_{mk}=0$ for all other $m$, the Eq.(\ref{3prime}) and (\ref{7.a}) imply that 
\begin{eqnarray}
\Lambda_k \simeq \frac{\tau_{0k}}{\vert T_{0k}\vert}~.
\label{6b}
\end{eqnarray}
The in-plane spin for the Eq.(\ref{6b}) is shown in Fig.\ref{spin-canting_a}. A nonzero $\tau_{0k}$ is yet another anomalous effect which may be caused by non-central potentials. For $m=1$,
\begin{eqnarray}
\tau_{1k}=\frac{2}{\hbar}
\int_{-\pi}^{\pi}\frac{d\phi}{2\pi}e^{-i6\phi} {\vec S}_{\vec k}.\hat{k}=\frac{2}{\hbar}\langle e^{-i6\phi} {\vec S}_{\vec k}.\hat{k}\rangle_\phi~.
\label{warping_vs_canting}
\end{eqnarray}
This gives symmetrically distributed canting of the spin with $2\pi/6$-period.   
The SCA data in Ref.\cite{gedik} is an example of Eq.(\ref{warping_vs_canting}) with a real $T_{0k}$. From Eq.(\ref{6}) and (\ref{warping_vs_canting}) we find 
\begin{eqnarray}
\Lambda_k(\phi) \simeq \frac{\tau_{1 k}}{T_{0k}}\sin (6\phi+\phi_0)
\label{6c}
\end{eqnarray} 
where $\phi_0$ is a reference angle fixed by the $T_{0k}$. The full agreement with the measured SCA in Ref.\cite{gedik} is obtained when $\phi_0=0$. Then, the $T_{0k}$ is real, and $\delta_{\vec k}$ vanishes along the high symmetry directions $\Gamma M, \Gamma K$ as shown in Fig.\ref{spin-canting_b}.

The Eq.'s(\ref{3prime})-(\ref{6c}) built the general connection between the $Im\{T_{\vec k}\}$ and the SCA. We now demonstrate that, the HW is responsible for the $Im\{T_{\vec k}\}$, establishing the full connection between the HW and the SCA. We begin with $V_{\vert{\vec k}-{\vec k}^\prime\vert}$ and assume that it has a typical energy scale $V_0$ much smaller than any energy scale in the SPP, i.e. $V_0 \ll \alpha k_F, \lambda k_F^3 \ll E_F$\cite{k_F}. The  off-diagonal SE at the Hartree-Fock level is $\Sigma_{xy\vec k}=-\sum_{\vec k^\prime}V_{\vert {\vec k}-{\vec k}^\prime\vert}\langle {\bf e}^\dagger_{{\vec k^\prime}\downarrow} {\bf e}_{{\vec k^\prime}\uparrow}\rangle$ calculated in the $\vert +\vec k^\prime \rangle$ state. For ${\cal H}^\prime$ in Eq.(\ref{4}) this is,
\begin{eqnarray}
\Sigma_{xy\vec k}&=& -\frac{1}{2}\sum_{{\vec k}^\prime} V_{\vert \vec k-\vec k^\prime\vert} \frac{g_{xy\vec k^\prime}+ \Sigma_{xy\vec k^\prime}}{\vert G_{\vec k^\prime}\vert}\,f_{\vec k^\prime}~~~~~~~~
\label{selfen}
\end{eqnarray}
with $f_{\vec k}=[1+exp(\beta E_{\vec k})]^{-1}$, $\beta$ as the inverse temperature and $E_{\vec k}=E_{0\vec k}+\vert G_{\vec k}\vert$. We consider a zero temperature formulation here, hence $f_{\vec k^\prime}=\Theta(\mu-E_{\vec k^\prime})$. The Eq.(\ref{selfen}) allows complex solutions in the form of Eq.(\ref{6}) and (\ref{6a}) when the HW is present. 
Since the knowledge about the interactions is limited, the exact solution of Eq.(\ref{selfen}) is not necessary. We simplify the r.h.s. of Eq.(\ref{selfen}) in order to extract a simple analytic result with the essential features captured. Using Eq.(\ref{6}) and remembering that $Im\{T_{\vec k}\}=t_{\vec k}\sin\Lambda_{\vec k}$, we write the imaginary part of Eq.(\ref{selfen}) as\cite{supp},  
\begin{eqnarray}
Im\{T_k(\phi)\}=-\frac{1}{2}\sum_{\vec k^\prime} v_{kk^\prime}(\phi^\prime) \sin\phi^\prime \,{\cal L}^-_{\phi^\prime}\frac{\alpha k^\prime}{\vert G_{k^\prime}(\phi)\vert}~~~~
\label{imaginary_tk_22}
\end{eqnarray}
where ${\cal L}_{\phi^\prime}^-$ is an operator defined by the difference of two angular translations as ${\cal L}_{\phi^\prime}^- h(\phi)=h(\phi+\phi^\prime)- h(\phi-\phi^\prime)$ for an arbitrary anisotropic function $h(\phi)$. The Eq.(\ref{imaginary_tk_22}) is a clear statement that $Im\{T_k(\phi)\}$ originates from the angular anisotropy in the effective in-plane spin orbit coupling $G_k(\phi)$. We now approximate the ${\cal L}_{\phi^\prime}^-$ with its leading term which is a linear derivative, ${\cal L}_{\phi^\prime}^-\simeq 2\phi^\prime \partial/\partial_\phi$.   
The $Im\{T_{\vec k}\}$ can then be found as\cite{supp}  
\begin{eqnarray}
Im\{T_{\vec k}\}=t_{\vec k}\sin\Lambda_{\vec k}=
\frac{\lambda^2}{\alpha^2}{\cal S}_k\,\sin(6\phi)
\label{imaginary_tk_4}
\end{eqnarray}
where ${\cal S}_k$ is some radial "structure factor". In Ref.\cite{gedik} a polynomial (or power)-like dependence of the spin canting angle $\delta_{\vec k}$ was observed as a function of the energy $E$. We therefore consider a simple power-law $V_{{\vec k}-{\vec k^\prime}}=V_0 (\vert{\vec k}-{\vec k^\prime}\vert\,d)^p$ where $d=\sqrt{\lambda/\alpha}\simeq 16 \AA$ for $Bi_2Se_3$. We than have\cite{supp} ${\cal S}_k\sim V_0 k^{4+p}$. The spin canting angle can then be inferred from Eq.'s (\ref{7.a}) and (\ref{imaginary_tk_4}) as
\begin{eqnarray}
\delta_{\vec k}\simeq V_0 \frac{d}{\alpha}(kd)^{3+p}\sin({6\phi})
\label{imaginary_tk_5}
\end{eqnarray} 
Eq.(\ref{imaginary_tk_5}) is our main result which completes the promised connection between the SCA and the HW. 

\begin{figure}[h]
\includegraphics [height=6.8cm,width=8.88cm]{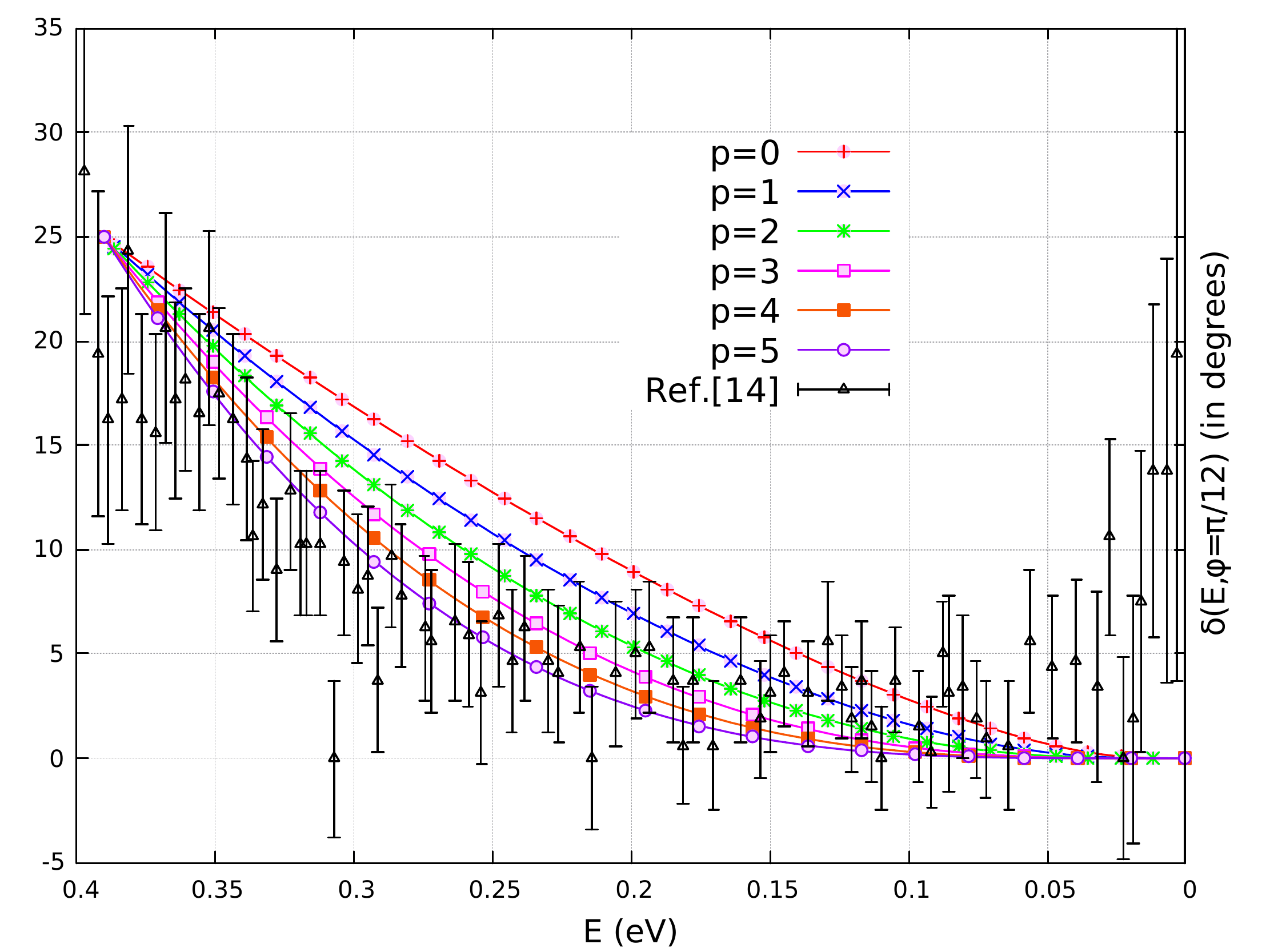}  
\vskip-0.3truecm
\caption{(Color online) Spin canting anomaly predicted by  Eq.(\ref{selfen}) using a power-like interaction for $V_{{\vec k} {\vec k^\prime}}=V_0 \vert {\vec k}- {\vec k^\prime}\vert^p$ with several $p$ values. We fixed the direction halfway between the $\Gamma M$ and $\Gamma K$ directions, where $\delta_{\vec k}$ is maximum. The data is reproduced from Ref.\cite{gedik} for comparison.}  
\label{canting_from_selfen}
\end{figure}

We report here two consequences of Eq.(\ref{imaginary_tk_5}) related to the existing experiments. Firstly, using the Eq.(\ref{6c}) in the Eq.(\ref{3prime}) the $x$ component of the spin is found to be 
\begin{eqnarray}
S_{x \vec k}=\frac{\hbar}{2} \sqrt{1-\frac{(G_z)^2}{\vert {\vec G}_{\vec k}\vert^2}}\sin(\phi+\delta_{\vec k})
\label{spin-canting_c}
\end{eqnarray} 
This equation, which was suggested as an empirical form of the experimental data in Ref.\cite{gedik,Y_Wang}, follows naturally from our theory. The second consequence is that, the power law provides a $k^{3+p}$ dependence to $\delta_{\vec k}$ as in Eq.(\ref{imaginary_tk_5}). At a fixed energy $E$, $k=k(E)$ and this implies a polynomial for $\delta_{\vec k}$ in low powers of $E$. Using the Fermi velocity $v_{BS}\simeq 3.5 eV \AA^{-1}$, the HW strength $\lambda_{BS}\simeq 128 eV\AA^3$ and the Fermi energy $E_{F(BS)}\simeq 0.3 eV$ for $Bi_2Se_3$ in Ref.\cite{gedik} and \cite{kuroda} we can estimate the effective interaction strength $V_0$ in Eq.(\ref{imaginary_tk_5}) as $1-5 meV$ for $p=3$. The corresponding $\delta_{\vec k}$ vs. $E$ is calculated numerically for $Bi_2Se_3$ in Fig.(\ref{canting_from_selfen}).  

The present theory successfully provides an understanding of the spin canting and an order of magnitude estimation for the interaction. However this picture is insufficient in understanding the experimental behaviour of $\delta_{\vec k}$ in low energies. To show this, we reproduced from the original experiment the $\delta_{\vec k}$ vs. $E$ data in Fig.(\ref{canting_from_selfen}). The  theoretical model clearly deviates from the reproduced data in the vicinity of the DP. Large error bars of the data in this regime also make a unique interpretation difficult. There may be several reasons for this behaviour. Firstly, there are technical difficulties in the momentum resolved measurements near the DP due to the vanishing ${\vec k}$-space area. Secondly, the experimental energy resolution may not allow a precise separation of the upper cone energetically from the lower one near the DP. If this is the case, $\delta_{\vec k}$ should have an asymmetric energy dependence around the DP (shown in the original reference \cite{gedik} and not in Fig.(\ref{canting_from_selfen}) here). We therefore rule this out. Finally, there may be increased quasiparticle scattering due to $q=0$ optical phonons. This can cause a broadening of the momentum distribution away from the sharp spin-momentum locking via the diagonal self-energies and influence the spin distribution via the non-diagonal ones. These two effects combine in a similar broadening of $\delta_{\vec k}$ and an increase in its mean, thus causing the discrepancy between the theory and the data in the low energies.   

The third scenario above is consistent with the behaviour of the dimensionless electron-optical phonon (e-ph) coupling $\lambda(E)$ at small $E$ as recently reported in the ab-initio analysis\cite{Heid}. The e-ph interaction has been heavily examined \cite{Zhu,Pan,A1g_Bi2Se3} as the dominant interaction mechanism at the room temperature and the $\lambda(E)$ studied both experimentally and theoretically varies in a large range from exceptionally weak\cite{Pan} to strong coupling\cite{Zhu} for branch specific phonons. The authors in Ref.\cite{Heid} have shown that the $\lambda(E)$ remarkably follows the topological surface DOS within the bulk band gap. The linear $E$ dependence is also consistent with the Eliashberg theory which provides a simple form $\lambda(E)\simeq 2 \rho(E) g^2_{e-ph}/\omega_0$ where $\rho(E)$ is the electron DOS, $g_{e-ph}$ is the microscopic e-ph coupling and $\omega_0$ is a typical optical phonon frequency. The behaviour of $\lambda(E)$ around the DP is exceptionally large and this is either due to the contamination of the lower Dirac cone or a high DOS near $E=0$. We find it remarkable that the energy dependence of the e-ph coupling in Ref.\cite{Heid} and the $\delta_{\vec k}$ data in Ref.\cite{gedik} are very similar. 

The quintic Dresselhaus SOC proposed in Ref.\cite{Basak} for explaining a similar SCA for $Bi_2Te_3$ introduces a new coupling constant which may also be relevant at a weaker scale in $Bi_2Se_3$. However, the lack of the experimental verification of the quintic Dresselhaus spin-orbit coupling as of this date, is an impediment for understanding the SCA. On the other hand, in the context of this work, the SCA is a natural consequence of the interactions which have been experimentally observed. Moreover, the interaction-based scenario formulated here is  a consequence of the hexagonal warping. New experiments with the scope of observing the effect of the HW on the non-diagonal SE will be highly illuminating in this direction.    

That the spin information in the measured photoelectrons' final state can be different from that in the original TSS depending on the polarization of light was theoretically\cite{Park_Louie} and experimentally\cite{Jozwiak_Barriga2}  shown to be inherent to the CD-ARPES technique. This raised criticism about the interpretation of the measurements in Ref.\cite{gedik}. This does not affect our conclusions here in that, the comparison of our theoretical model with the relevant results in Ref.\cite{gedik} is solely based on the six-fold canting symmetry in the spin configuration of the photoemitted electrons. Since the light sources used are polarized linearly or (left/right) circularly, the basic symmetries on the initial state are naturally reflected on the final state\cite{Park_Louie}. Hence, the observed six-fold SCA is an intrinsically real property of the initial state.  

\section{Acknowledgement}
The author thanks Nikesh Koirala, Nuh Gedik and Liang Fu for useful discussions and the MIT for hospitality during the visit where this work was initiated.  
\renewcommand{\theequation}{A-\arabic{equation}}
\renewcommand{\thesection}{S\arabic{section}}
\renewcommand{\thetable}{S\arabic{table}}  
\renewcommand{\thefigure}{S\arabic{figure}}
\setcounter{equation}{0}  
\section{Appendix}
In order to obtain the Eq.(\ref{imaginary_tk_22}), we use Eq.(\ref{6}) in (\ref{selfen}) and obtain 
\begin{equation}
T_{\vec k}= -\frac{1}{2}\sum_{{\vec k}^\prime} V_{\vert \vec k-\vec k^\prime\vert} e^{i(\phi_S^\prime-\phi_S)} \frac{\alpha k^\prime+ T_{\vec k^\prime}}{\vert G_{\vec k^\prime}\vert}\,f^{+}_{\vec k^\prime}~~~~~~~~
\label{selfen_supp}
\end{equation}
After some algebra and a shift of variables $\phi^\prime \to \phi^\prime+\phi$ on the r.h.s., of the Eq.(\ref{selfen_supp}) here, the imaginary part of $Im\{T_{\vec k}\}$ can be found as, 

\begin{eqnarray} 
&~&Im\{T_{\vec k}\}=-\frac{1}{2}\sum_{{\vec k}^\prime} v_{kk^\prime}(\phi^\prime) e^{-i\phi^\prime} \nonumber \\ 
&\times&\Biggl\{\Bigl[\frac{\alpha k^\prime+ Re\{T_{k^\prime}(\phi^\prime+\phi)\}}{\vert G_{k^\prime}(\phi^\prime+\phi)\vert}-\frac{\alpha k^\prime+ Re\{T_{k^\prime}(-\phi^\prime+\phi)\}}{\vert G_{k^\prime}(-\phi^\prime+\phi)\vert}\Bigr] \nonumber \\
&+&i\Bigl[\frac{Im\{T_{k^\prime}(\phi^\prime+\phi)\}}{\vert G_{k^\prime}(\phi^\prime+\phi)\vert}+\frac{Im\{T_{k^\prime}(-\phi^\prime+\phi)\}}{\vert G_{k^\prime}(-\phi^\prime+\phi)\vert}\Bigr]\Biggl\}
\label{selfen_supp_im}
\end{eqnarray}
where $v_{kk^\prime}(\phi^\prime)$ is essentially the same as $V_{\vert \vec k-\vec k^\prime\vert}$ with the replacement $\phi^\prime \to \phi^\prime+\phi$. We also introduced a slightly new notation $G_{k^\prime}(\phi)$ and $T_{k^\prime}(\phi)$ separating the radial and the azymuthal angular dependences in order to independently analyze the shifted angular dependence of $G_{\vec k}$ and $T_{\vec k}$. We now ignore in the r.h.s. of Eq.(\ref{selfen_supp_im}), the $Re\{T_{\vec k^\prime}\}$ as compared to the $\alpha k^\prime$, and define two new operators ${\cal L}_{\phi^\prime}^\pm$ by their action on an arbitrary function $h(\phi)$  
\begin{eqnarray}
{\cal L}_{\phi^\prime}^\pm \, h(\phi)=h(\phi+\phi^\prime) \pm h(\phi-\phi^\prime)
\end{eqnarray}
and write the Eq.(\ref{selfen_supp_im}) in a more suggestive form as 
\begin{eqnarray}
Im\{T_k(\phi)\}&=&-\frac{1}{2}\sum_{\vec k^\prime} v_{kk^\prime}(\phi^\prime) \,\nonumber \\
\times \Biggl\{\sin\phi^\prime \,{\cal L}^-_{\phi^\prime}&\Bigl[&\frac{\alpha k^\prime+Re\{T_{k^\prime}(\phi)\}}{\vert {\vec G}_{k^\prime}(\phi)\vert}\Bigr] \nonumber \\
&+&\cos\phi^\prime {\cal L}^+_{\phi^\prime}\Bigl[\frac{Im\{T_{k^\prime}(\phi)\}}{\vert {\vec G}_{k^\prime}(\phi)\vert}\Bigr]\Biggr\}
\label{imaginary_tk_2}
\end{eqnarray}  
The Eq.(\ref{imaginary_tk_2}) must be solved self-consistently for $T_{\vec k}$ with a known interaction $v_{k k^\prime}(\phi^\prime)$. We claim that a general result not depending on the details of the interactions is more illustrative than the full self-consistent solution of Eq.(\ref{imaginary_tk_2}). Firstly, if the interaction is sufficiently weaker than the SOC, we can ignore the $T_{\vec k}$ dependent terms on the r.h.s. of Eq.(\ref{imaginary_tk_2}). Secondly, the ${\cal L}^+_{\phi^\prime}$ dependent term is approximately a renormalizion of the $Im\{T_k(\phi)\}$ on the l.h.s. and brings a $k$-dependent overall factor. This detail can be neglected for a simple result which retains only the most essential properties of the solution. We are then left with the ${\cal L}^-_{\phi^\prime}$ dependent term only. The result is,
\begin{equation}
Im\{T_k(\phi)\}=-\frac{1}{2}\sum_{\vec k^\prime} v_{kk^\prime}(\phi^\prime) \sin\phi^\prime \,{\cal L}^-_{\phi^\prime}\Bigl[\frac{\alpha k^\prime}{\vert G_{k^\prime}(\phi)\vert}\Bigr] ~~
\label{imtk}
\end{equation}
which is the Eq.(\ref{imaginary_tk_22}). This indicates that a nonzero $Im\{T_k(\phi)\}$ is the result of the anisotropy in $G_{k^\prime}(\phi)$. Since the source of anisotropy is the hexagonal warping in the context of this work, Eq.(\ref{imtk}) builts the connection between the spin canting anomaly and the hexagonal warping as explained below Eq.(\ref{imaginary_tk_22}). The angular translation operators in ${\cal L}^-_{\phi^\prime}$ act on the HW part in $\vert G_{k^\prime}(\phi)\vert$. Replacing the ${\cal L}^-_{\phi^\prime}$ with its leading term $2\phi^\prime \partial/\partial \phi$, and defining 
\begin{equation}
{\cal S}_k=\frac{3}{2}\sum_{\vec k^\prime}\phi^\prime \sin\phi^\prime {k^\prime}^4 v_{k k^\prime}(\phi^\prime) 
\label{Sk}
\end{equation} 
The Eq.(\ref{Sk}) is then used in the Eq.(\ref{imaginary_tk_4}).
 


\begin{thebibliography}{99} 
\bibitem{TI_general_intro} C.L. Kane and E.J. Mele, Phys. Rev. Lett., {\bf 95}, 226801 (2005); ibid 146802 (2005); L. Fu, C.L. Kane and E.J. Mele, Phys. Rev. Lett. {\bf 98} 106803 (2007); J.E. Moore and L. Balents, Phys. Rev. {\bf 75}, 121306(R) (2007); Bernevig, B. A., T. A. Hughes, and S.C. Zhang, Science {\bf 314}, 1757 (2006); L. Fu and C.L. Kane, Phys. Rev. {\bf B76}, 045302 (2007); D. Hsieh, D. Qian, L. Wray, Y. Xia, Y.S. Hor, R.J. Cava and M.Z. Hasan, Nature (London) {\bf 452}, 970 (2008)
\bibitem{topo_protect} J. E. Moore, Nature {\bf 464}, 194 (2010); M. Z. Hasan and C, L. Kane,Rev. Mod. Phys.82, 3045 (2010); X.-L. Qi and S.-C. Zhang,Phys. Today63, 33 (2010).
\bibitem{Park} S. R. Park, W. S. Jung, Chul Kim, D. J. Song, C. Kim, S. Kimura, K. D. Lee and N. Hur, Phys. Rev. {\bf B81}, 041405 (2010).
\bibitem{Chen} Chaoyu Chen, Zhuojin Xie, Ya Feng et al. Nature Sci. Rep., {\bf 3}, 2411 (2013). 
\bibitem{Zhu} Richard C. Hatch, Marco Bianchi, Dandan Guan, Shining Bao, Jianli Mi, Bo Brummerstedt Iversen, Louis Nilsson, Liv Hornekaer, and Philip Hofmann, Phys. Rev. {\bf B83}, 241303 (2011); X.Zhu, L. Santos, C.Howard, R. Sankar, F.C. Chou, C. Chamon and M.El-Batanouny, Phys. Rev. Lett. {\bf 108}, 185501 (2012); A. D. LaForge, A. Frenzel, B. C. Pursley, Tao Lin, Xinfei Liu, Jing Shi, and D. N. Basov, Phys. Rev. {\bf B 81}, 125120 (2010). 
\bibitem{Pan} Z-H. Pan, A.V. Fedorov, D. Gardner, Y.S. Lee, S. Chu and T. Valla, Phys. Rev. Lett. {\bf 108}, 187001 (2012). 
\bibitem{Valla} T. Valla, Z.-H. Pan, D. Gardner, Y.S. Lee and S. Chu, Phys. Rev. Lett. {\bf 108}, 117601 (2012).
\bibitem{Barriga} J. Sanchez-Barriga, M. R. Scholz, E. Golias, E. Rienks, D. Marchenko, A. Varykhalov, L. V. Yashina and O. Rader, Phys. Rev. {\bf B90}, 195413 (2014).
\bibitem{Heid} R. Heid, I. Yu. Sklyadneva, and E. V. Chulkov, Sci. Rep. {\bf 7}, 1095 (2017)
\bibitem{BiSe_exp} Y. Xia, D. Qian, D. Hsieh, L. Wray, A. Pal, A. Bansil, D. Grauer, Y. S. Hor, R. J. Cava, and M. Z. Hasan, Nature Phys. {\bf 5}, 398 (2009)
\bibitem{BiTe} Y. L. Chen, J. G. Analytis, J.-H. Chu, Z. K. Liu, S.-K. Mo, X. L. Qi, H. J. Zhang, D. H. Lu, X. Dai, Z. Fang, S. C. Zhang, I. R. Fisher, Z. Hussain, Z.-X. Shen, Science {\bf 325}, 178 (2009); D. Hsieh, Y. Xia, D. Qian, L. Wray, F. Meier, J. H. Dil, J. Osterwalder, L. Patthey, A. V. Fedorov, H. Lin, A. Bansil, D. Grauer, Y. S. Hor, R. J. Cava, and M. Z. Hasan, Phys. Rev. Lett. {\bf 103}, 146401 (2009); Z. Alpichshev, J. G. Analytis, J.-H. Chu, I. R. Fisher, Y. L. Chen, Z. X. Shen, A. Fang, and A. Kapitulnik, Phys. Rev. Lett. {\bf 104}, 016401 (2010).
\bibitem{BiSe_ARPES} E. Lahoud,  E. Maniv, M. S. Petrushevsky, M. Naamneh, A. Ribak, S. Wiedmann, L. Petaccia, Z. Salman, K. B. Chashka, Y. Dagan, and A. Kanigel, Phys. Rev. B 88, 195107 (2013).
\bibitem{Fu_1} L. Fu, Phys. Rev. Lett. {\bf 103},266801 (2009); L. Fu, Phys. Rev. {\bf B 90}, 100509(R) (2014). 
\bibitem{gedik} Y. H. Wang, D. Hsieh, D. Pilon, L. Fu, D. R. Gardner, Y. S. Lee, and N. Gedik, Phys. Rev. Lett. 107, 207602 (2011).
\bibitem{criticism} The discussions about the CD-ARPES experiments and their criticism have been made in the final part of this manuscript.  
\bibitem{DOI} Fabian Meier, Hugo Dil, Jorge Lobo-Checa, Luc Patthey, and Jurg Osterwalder, Phys. Rev. {\bf B 77}, 165431 (2008) (DOI 10.1103/PhysRevB.79.089902)
\bibitem{Basak} Susmita Basak, Hsin Lin, L.A. Wray, S.-Y. Xu, L. Fu, M. Z. Hasan, A. Bansil Phys. Rev. {\bf B 84}, 121401 (2011).
\bibitem{qp_interference} S. Souma, K. Kosaka, T. Sato, M. Komatsu, A. Takayama, T. Takahashi, M. Kriener, Kouji Segawa, and Yoichi Ando, Phys. Rev. Lett. {\bf 106}, 216803 (2011).
\bibitem{k_F} The anisotropy in the Fermi momentum $k_F$ can be ignored here.

\bibitem{supp} See the Appendix
\bibitem{Y_Wang} Yihua Wang, {\it Laser Based Angle Resolved Photoemission Spectroscopy of Topological Insulators}, Harvard University Doctoral Dissertation (2012). 

\bibitem{kuroda} K. Kuroda, M. Arita, K. Miyamoto, M. Ye, J. Jiang, A. Kimura, E. E. Krasovskii, E. V. Chulkov, H. Iwasawa, T. Okuda, K. Shimada, Y. Ueda, H. Namatame and M. Taniguchi Phys. Rev. Lett. {\bf 105}, 076802 (2010).
\bibitem{A1g_Bi2Se3} J. A. Sobota, S.-L. Yang, D. Leuenberger, A. F. Kemper, J. G. Analytis, I. R. Fisher, P. S. Kirchmann, T. P. Devereaux, and Z.-X. Shen, Phys. Rev. Lett. {\bf 113}, 157401 (2014).
\bibitem{Park_Louie} C.H. Park and S.G. Louie, Phys. Rev. Lett. {\bf 109}, 097601 (2012)
\bibitem{Jozwiak_Barriga2} C. Jozwiak, C.-H. Park, K. Gotlieb, C. Hwang, D.-H. Lee, S. G. Louie, J.D. Denlinger, C.R. Rotundu, R.J. Birgeneau, Z. Hussain and A. Lanzara, Nat. Phys. {\bf 9}, 293 (2013); J. Sanchez-Barriga, A. Varykhalov, J. Braun, S.-Y. Xu, N. Alidoust, O. Kornilov, J. Minar, K. Hummer, G. Springholz, G. Bauer, R. Schumann, L. V. Yashina, H. Ebert, M. Z. Hasan, and O. Rader, Phys. Rev. {\bf X4}, 011046 (2014).
\end{thebibliography}
\end{document}